\documentclass[aps,prd,twocolumn,showpacs,showkeys,amsmath,amssymb,nofootinbib]{revtex4}
\usepackage{amsfonts}
\usepackage{graphicx}
\usepackage{bm}
\usepackage{graphics,booktabs}
\usepackage{longtable,lscape}
\usepackage{txfonts}
\usepackage{overpic}
\usepackage{amssymb}
\usepackage{epsfig}
\usepackage{feynmf}
\usepackage{epstopdf}
\usepackage{slashed}
\usepackage{multirow}
\usepackage{color}
\usepackage{ulem}
\usepackage[colorlinks,citecolor=blue,anchorcolor=red,menucolor=red,linkcolor=red,filecolor=red,runcolor=red,urlcolor=blue,frenchlinks=red]{hyperref}

\makeatletter
\newcommand{\figcaption}{\def\@captype{figure}\caption}
\newcommand{\tabcaption}{\def\@captype{table}\caption}

\newcommand{\Rmnum}[1]{\expandafter\@slowromancap\romannumeral #1@}
\def\hlinewd#1{%
  \noalign{\ifnum0=`}\fi\hrule \@height #1 \futurelet
   \reserved@a\@xhline}
\makeatother

\def\qq{\langle\bar qq\rangle}
\def\ss{\langle \bar ss\rangle}

\def\GGb{\langle g_s^2GG\rangle}
\def\qGqa{\langle g_s\bar q\sigma Gq\rangle}

\def\sGsa{\langle g_s\bar s\sigma Gs\rangle}

\def\f(s){[(\alpha+\beta)m^2-\alpha\beta s]}

\begin{document}

\title{$X_0(2900)$ and $X_1(2900)$: hadronic molecules or compact tetraquarks}

\author{Hua-Xing Chen$^1$}
\email{hxchen@buaa.edu.cn}
\author{Wei Chen$^2$}
\email{chenwei29@mail.sysu.edu.cn}
\author{Rui-Rui Dong$^3$}
\author{Niu Su$^3$}
\affiliation{
$^1$School of Physics, Southeast University, Nanjing 210094, China\\
$^2$School of Physics, Sun Yat-Sen University, Guangzhou 510275, China\\
$^3$School of Physics, Beihang University, Beijing 100191, China
}

\begin{abstract}
Very recently the LHCb Collaboration reported their observation of the first two fully open-flavor tetraquark states, the $X_0(2900)$ of $J^P = 0^+$ and the $X_1(2900)$ of $J^P = 1^-$. We study their possible interpretations using the method of QCD sum rules, paying special attention to an interesting feature of this experiment that the higher resonance $X_1(2900)$ has a width significantly larger than the lower one $X_0(2900)$. Our results suggest that the $X_0(2900)$ can be interpreted as the $S$-wave $D^{*-}K^{*+}$ molecule state of $J^P = 0^+$, and the $X_1(2900)$ can be interpreted as the $P$-wave $\bar c \bar s u d$ compact tetraquark state of $J^P = 1^-$. Mass predictions of their bottom partners are also given.
\end{abstract}

\maketitle

$\\$
{\it Introduction.}-----
Very recently the LHCb Collaboration studied the $B^+ \to D^+ D^- K^+$ decay~\cite{lhcb}. They observed an exotic peak in the $D^- K^+$ invariant mass spectrum with a significance overwhelmingly larger than $5\sigma$. They used two Breit-Wigner resonances with spin-0 and spin-1 to fit this peak, whose parameters are extracted to be
\begin{eqnarray}
\label{experiment}
     X_0(2900) &:&    M= 2.866 \pm 0.007 \pm 0.002  \mbox{ GeV} \, ,
\\ \nonumber            &&    \Gamma=    57 \pm 12 \pm 4  \mbox{ MeV} \, ,
\\ \nonumber     X_1(2900)&:&    M= 2.904 \pm 0.005 \pm 0.001  \mbox{ GeV} \, ,
\\ \nonumber            &&    \Gamma=    110 \pm 11 \pm 4  \mbox{ MeV} \, .
\end{eqnarray}
These two resonances were observed in the $D^- K^+$ final state, implying: a) the possible spin-parity quantum numbers of $X_0(2900)$ and $X_1(2900)$ are $J^P = 0^+$ and $1^-$, respectively; and b) their quark contents are $\bar c \bar s u d$, so they are exotic hadrons with valence quarks of four different flavors. The latter point makes this experiment very interesting, because this is the first time observing such exotic hadrons~\cite{pdg}. Their theoretical and experimental studies will significantly improve our understanding on the non-perturbative behaviors of the strong interaction at the low energy region~\cite{Chen:2016qju,Lebed:2016hpi,Esposito:2016noz,Guo:2017jvc,Ali:2017jda,Olsen:2017bmm,Karliner:2017qhf,Liu:2019zoy,Brambilla:2019esw}. Note that the D0 Collaboration reported the evidence of another narrow structure $X(5568)$ in 2016~\cite{D0:2016mwd,Abazov:2017poh}, which is also expected to be composed of four different flavors, while its existence was not confirmed in the following LHCb, CMS, CDF, and ATLAS experiments~\cite{Aaij:2016iev,Sirunyan:2017ofq,Aaltonen:2017voc,Aaboud:2018hgx}.

This experiment has an interesting feature that the higher resonance $X_1(2900)$ of $J^P = 1^-$ has a width significantly larger than the lower one $X_0(2900)$ of $J^P = 0^+$, although the $X_1(2900) \to D^-K^+$ decay is $P$-wave and the $X_0(2900) \to D^-K^+$ decay is $S$-wave. Paying special attention to this feature, in this letter we study their possible interpretations using the QCD sum rules, a powerful and successful non-perturbative method, which have been widely applied to study various exotic hadrons~\cite{Shifman:1978bx,Reinders:1984sr}.

The lower resonance $X_0(2900)$ of $J^P = 0^+$ has the mass $M_{X_0} = 2866.3 \pm 6.5 \pm 2.0$~MeV, that is about $36\pm7$~MeV below the $D^{*-} K^{*+}$ threshold~\cite{pdg}. In this letter we study it using the meson-meson tetraquark current of $J^P = 0^+$, which well couples to the $D^{*-}K^{*+}$ molecular state of $J^P = 0^+$. Through the QCD sum rule method, we extract the hadron mass to be $m_{X_0,\, 0^+} = 2.87^{+0.19}_{-0.14}$~GeV. This suggests a possible $J^P = 0^+$ $D^{*-}K^{*+}$ molecule interpretation for the $X_0(2900)$. In this picture the width of the $X_0(2900)$ can be naturally explained: it is mainly contributed by the width of the $K^{*+}$ meson, that is $\Gamma_{K^{*+}} = 50.8 \pm 0.9$~MeV~\cite{pdg}; this further suggests that the $X_0(2900)$ mainly decays into the $D^{*-} K^{+} \pi^{0}/D^{*-} K^{0} \pi^{+}$ three-body final states, and we propose to investigate these channels to confirm the $X_0(2900)$.

The higher resonance $X_1(2900)$ of $J^P = 1^-$ has the mass $M_{X_1} = 2904.1 \pm 4.8 \pm 1.3$~MeV, so it is about $38$~MeV above the $X_0(2900)$. This seems to suggest the $X_1(2900)$ to be the excited $D^{*-}K^{*+}$ molecular state of $J^P = 1^-$, but there arises a question: why it has a width significantly larger than the $X_0(2900)$, given its decay to $D^-K^+$ is $P$-wave. To find a consistent explanation for both $X_0(2900)$ and $X_1(2900)$, in this letter we also study the diquark-antidiquark tetraquark current of $J^P = 1^-$. We find one such current, from which the hadron mass is extracted to be $m_{X_1,\, 1^-}= 2.93^{+0.13}_{-0.11}$~GeV. This suggests a possible $J^P = 1^-$ $\bar c \bar s u d$ compact tetraquark interpretation for the $X_1(2900)$. In this picture the width of the $X_1(2900)$ is also reasonable, e.g., it is comparable to the $X(6900)$, a good $P$-wave $\bar c \bar c c c$ compact tetraquark candidate~\cite{Chen:2016jxd,Chen:2020xwe}, whose width is at the level of 100~MeV~\cite{Aaij:2020fnh}. Note that there also exist some other possible assignments for the $X(6900)$, e.g., see Refs.~\cite{Liu:2020orv,He:2020btl}.


Besides clarifying properties of the $X_0(2900)$ and $X_1(2900)$, in this letter we further give theoretical predictions of their bottom partners: the mass of the $S$-wave $B^{*0}K^{*+}$ molecule state with $J^P = 0^+$ is calculated to be $m_{X_{b0},\, 0^+}=6.16^{+0.22}_{-0.25}$~GeV, and the mass of the $P$-wave $\bar b \bar s u d$ compact tetraquark state with $J^P = 1^-$ is calculated to be $m_{X_{b1},\, 1^-}=6.27^{+0.21}_{-0.22}$~GeV. As the first two fully open-flavor tetraquarks, the observation of the $X_0(2900)$ and $X_1(2900)$ by LHCb~\cite{lhcb} has opened a new window for studying exotic hadrons. Further experimental and theoretical studies are crucially needed to understand them. The present study provides valuable information for future experimental exploration of these fully open-flavor tetraquarks.

$\\$
{\it Interpretation of the $X_0(2900)$ with $J^P = 0^+$.}-----
The peak observed by LHCb is at about 2.9~GeV, very close to the $D^{*-} K^{*+}$ threshold at 2902~MeV. In this section we investigate the relevant meson-meson tetraquark interpolating current with the spin-parity quantum number $J^P = 0^+$:
\begin{equation}
\eta(x) = \bar c_a(x) \gamma_\mu d_a(x) ~ \bar s_b(x) \gamma^\mu u_b(x) \, .
\label{def:eta}
\end{equation}
Here the subscripts $a$ and $b$ are color indices; $u,d,s,c$ represent the {\it up}, {\it down}, {\it strange}, and {\it charm} quarks, respectively. In this study we only consider local currents, so the four quarks/antiquarks are at the same location $x$, which parameter will be omitted in the following discussions.

The current $\eta$ can be naturally separated into two color-singlet components $\bar c_a \gamma_\mu d_a$ and $\bar s_b \gamma_\mu u_b$. These are two standard meson operators, well coupling to the $D^{*-}$ and $K^{*+}$ mesons respectively. Accordingly, $\eta$ would well couple to the $S$-wave $D^{*-}K^{*+}$ molecular state of $J^P = 0^+$, if such a state exists. Assuming the coupling to be
\begin{equation}
\langle 0 | \eta | X_0 \rangle = f_{X_0} \, ,
\label{eq:coupling1}
\end{equation}
we use the current $\eta$ to perform QCD sum rule analyses. To do this we need to calculate the two-point correlation function,
\begin{equation}
\Pi(p^{2}) = i\int d^4x ~ e^{ip\cdot x} ~ \langle0| {\bf T} [\eta(x)\eta^{\dag}(0)]|0\rangle\, ,
\label{equ:Pi1}
\end{equation}
at both hadron and quark-gluon levels.

At the hadron level we describe $\Pi(p^{2})$ in the form of the dispersion relation
\begin{equation}
\Pi(p^2) = \frac{(p^2)^N}{\pi}\int_{<}^{\infty}\frac{\mbox{Im}\Pi(s)}{s^N(s-p^2-i\epsilon)}ds + \sum_{n=0}^{N-1}b_n(p^2)^n \, ,
\label{eq:disper}
\end{equation}
where $b_n$ are subtraction constants, and can be removed later by performing the Borel transformation; $<$ is used to denote the physical threshold. We further define the imaginary part of the correlation function as the spectral density $\rho(s)\equiv{\mbox{Im}\Pi(s)}/{\pi}$, and write it as a sum over $\delta$ functions by inserting intermediate hadrons $|n\rangle$:
\begin{eqnarray}
\nonumber \rho(s) &=& \sum_n\delta(s-m_n^2)\langle0| \eta |n\rangle\langle n| \eta^{\dagger}|0\rangle+\mbox{continuum}
\\ &=& f_{X_0}^2\delta( s - m_{X_0}^2 )+ \mbox{continuum} \, .
\label{eq:imaginary}
\end{eqnarray}
The current $\eta$ can couple to all intermediate hadrons with the same quantum numbers. In QCD sum rule studies we usually investigate only the lowest-lying resonance, that is $|X_0\rangle$ in this case, with $m_{X_0}$ and $f_{X_0}$ to be its mass and decay constant respectively.

At the quark-gluon level, we evaluate $\Pi(p^{2})$ using the method of operator product expansion (OPE). In this study we calculate OPEs up to the $D({\rm imension}) = 8$ terms, including the perturbative term, the charm and strange quark masses, the quark condensate, the gluon condensate, the quark-gluon mixed condensate, and their combinations. The full expressions are too length to be shown here.

After performing the Borel transformation to pick out the lowest-lying state and remove the unknown subtraction constants $b_n$ in Eq.~(\ref{eq:disper}), we can compare $\Pi(p^2)$ at both phenomenological and OPE sides, and obtain the following QCD sum rule equation via the quark-hadron duality
\begin{equation}
\mathcal{L}_{k}\left(s_0,M_B^2\right)=f_X^2m_X^{2k}e^{-m_X^2/M_B^2}=\int_{<}^{s_0}dse^{-s/M_B^2}\rho(s)s^k\, ,
\label{eq:sumrule}
\end{equation}
where $s_0$ and $M_B$ are the threshold value and Borel mass, respectively.
The hadron mass is then calculated as
\begin{equation}
m_{X_0}\left(s_0, M_B^2\right)=\sqrt{{\mathcal{L}_{1}\left(s_0,M_B^2\right)\over\mathcal{L}_{0}\left(s_0, M_B^2\right)}}\, .
\label{eq:mass}
\end{equation}

The hadron mass $m_{X_0}$ in Eq.~(\ref{eq:mass}) is a function of $s_0$ and $M_B$, which are two essential parameters in QCD sum rule studies. We use two criteria to fix $M_B$: a) its lower bound is constrained by the OPE convergence, requiring the contribution of the highest terms ($D=8$ terms) to be less than $20\%$ of the total; b) its upper bound is constrained by the pole contribution (PC):
\begin{equation}
\text{PC}(s_0, M_B^2)=\frac{\mathcal{L}_{0}\left(s_0, M_B^2\right)}{\mathcal{L}_{0}\left(\infty, M_B^2\right)} \ge 30 \% \, .
\label{eq:pole}
\end{equation}
Altogether, we obtain a Borel window for a definite value of $s_0$.
Besides, the hadron mass $m_{X_0}$ should neither depend significantly on the unphysical parameter $M_B$, nor depend strongly on the threshold value $s_0$ in a reasonable region above $m_{X_0}$. This results in the stability criterion.

%
\begin{figure*}[hbt]
\begin{center}
\includegraphics[width=0.45\textwidth]{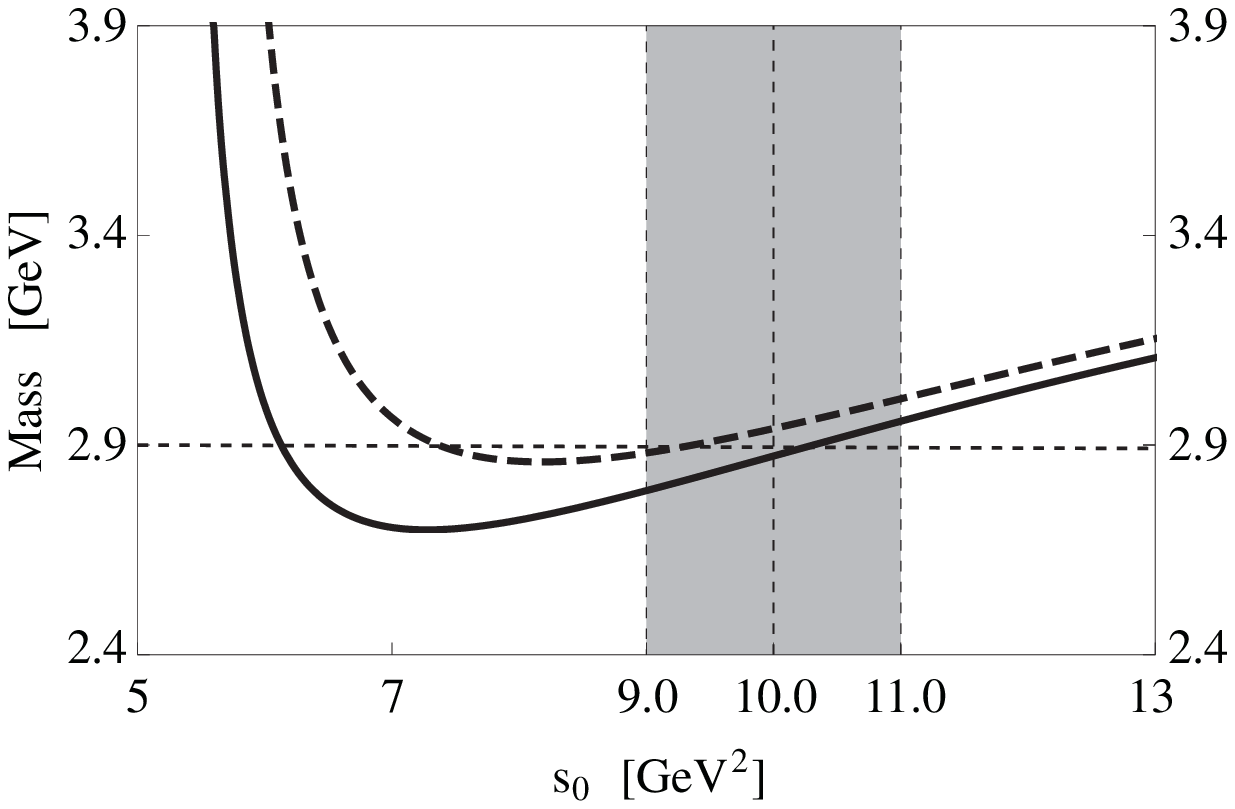}
~~~~~
\includegraphics[width=0.45\textwidth]{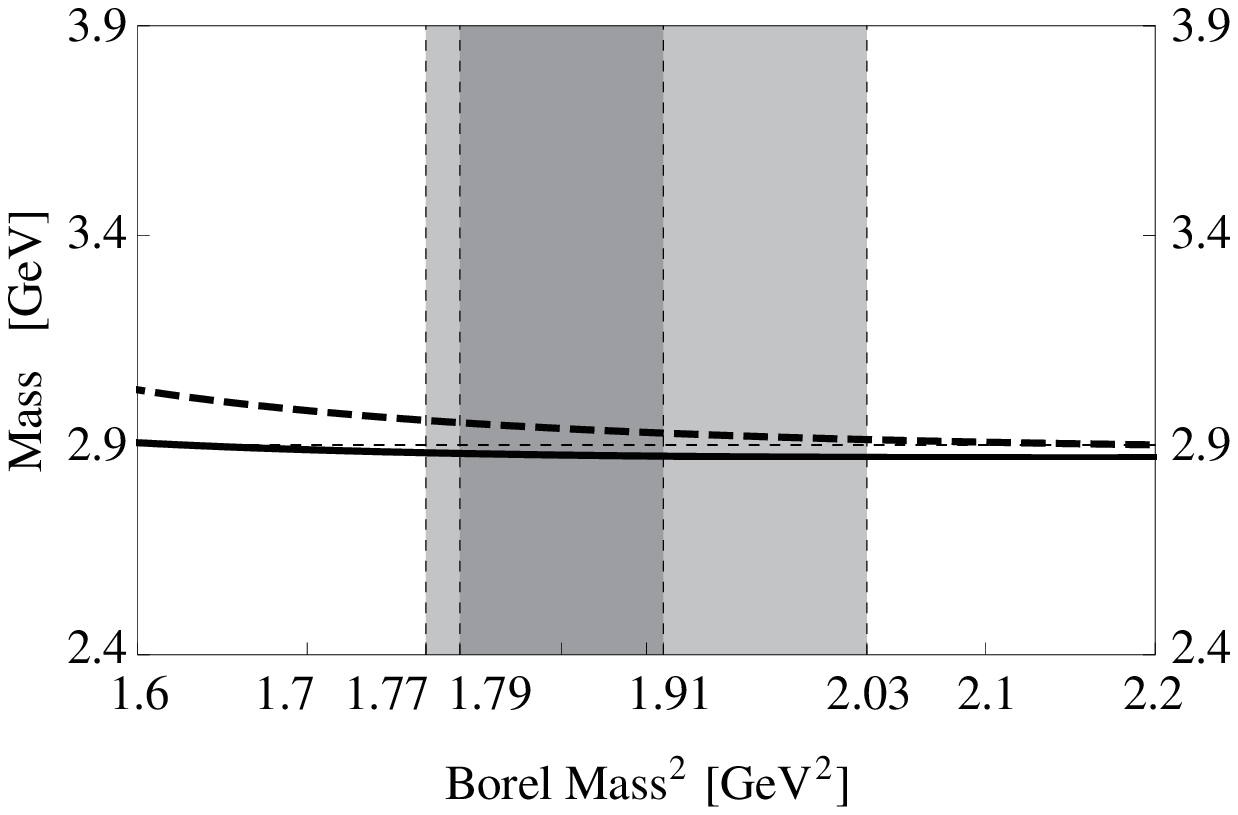}
\caption{
Masses calculated using the currents $\eta$ of $J^P=0^+$ (solid curves) and $\xi_\mu$ of $J^P=1^-$ (dashed curves), as functions of the threshold value $s_0$ (left) and the Borel mass $M_B$ (right). In the left panel the solid/dashed curves are obtained by setting $M_B^2 = 1.90/1.85$ GeV$^2$, respectively.
In the right panel the solid/dashed curves are both obtained by setting $s_0 = 10.0$ GeV$^2$, and the Borel windows for $\eta$ and $\xi_\mu$ are $1.77$ GeV$^2$ $\leq
M_B^2\leq 2.03$ GeV$^2$ and $1.79$ GeV$^2$ $\leq M_B^2\leq 1.91$ GeV$^2$, respectively.}
\label{fig:mass}
\end{center}
\end{figure*}
%

For the current $\eta$ of $J^P=0^+$, we show the variation of the
extracted mass $m_{X_0}$ with respect to the threshold value $s_0$ in the left panel of Fig.~\ref{fig:mass}. We show it using the solid curve,
and find that the mass dependence is moderate for
$9.0$ GeV$^2$ $\leq s_0\leq 11.0$ GeV$^2$, which is thus a reasonable working region.
Accordingly, we fix $s_0=10.0$ GeV$^2$, and then extract the Borel window to be
$1.77$ GeV$^2$ $\leq M_B^2\leq 2.03$ GeV$^2$ using the two criteria discussed above.

Within the above parameter regions, we show the variation of $m_{X_0}$
with respect to the Borel mass $M_B$ in the right panel of Fig.~\ref{fig:mass}. We show it still using the solid curve, and find it very stable inside the Borel window $1.77$ GeV$^2$ $\leq M_B^2\leq 2.03$ GeV$^2$, where the hadron mass is evaluated to be
\begin{equation}
m_{X_0,\, 0^+}=2.87^{+0.19}_{-0.14} \mbox{ GeV}\, .
\label{result:mass0}
\end{equation}
Here the uncertainty comes from the Borel mass $M_B$, the threshold value $s_0$, and various quark and gluon parameters. This value is in good agreement with the
mass of the $X_0(2900)$ measured by LHCb~\cite{lhcb}, suggesting a possible $J^P = 0^+$ $D^{*-}K^{*+}$ molecule interpretation for this exotic state. Employing the same formalism, we further study its bottom partner, the $S$-wave $B^{*0}K^{*+}$ molecule state of $J^P = 0^+$, and extract its mass to be
\begin{equation}
m_{X_{b0},\, 0^+}=6.16^{+0.22}_{-0.25} \mbox{ GeV}\, .
\label{result:massb0}
\end{equation}

In the above calculation we have adopted the following values for various quark and gluon parameters~\cite{Yang:1993bp,Narison:2002pw,Gimenez:2005nt,Jamin:2002ev,Ioffe:2002be,Ovchinnikov:1988gk,Ellis:1996xc,pdg}:
\begin{equation}
\begin{split}
& m_s(2\,\text{GeV}) = 95^{+9}_{-3} \text{ MeV} \, ,
\\ &
m_c(m_c) = 1.275^{+0.025}_{-0.035} \mbox{ GeV}   \, ,
\\ &
m_b(m_b) = 4.18^{+0.04}_{-0.03} \mbox{ GeV}   \, ,
\\&
\qq=-(0.24\pm0.01)^3 \mbox{ GeV}^3 \, ,
\\&
\ss=(0.8\pm0.1)~\qq \, ,
\\ &
\qGqa=-M_0^2~\qq \, ,
\\ &
\sGsa=-M_0^2~\ss \, ,
\\ &
M_0^2=(0.8\pm0.2)\text{ GeV}^2 \, ,
\\ &
\GGb= 0.48\pm0.14  \text{ GeV}^4 \, ,
\label{parameters}
\end{split}
\end{equation}
where we have used the running $charm$ and $bottom$ quark masses in the $\overline{\rm MS}$ scheme.

$\\$
{\it Interpretation of the $X_1(2900)$ with $J^P = 1^-$.}-----
In this section we investigate the diquark-antidiquark tetraquark interpolating currents with the spin-parity quantum number $J^P = 1^-$. Given the quark content to be $\bar c \bar s u d$, one can construct as many as sixteen independent diquark-antidiquark currents. We can generally write them as
\begin{equation}
\xi_\pm = \bar c_a \Gamma_1 C \bar s_b^T \left( u_a^T C \Gamma_2 d_b \pm u_b^T C \Gamma_2 d_a \right) \, ,
\end{equation}
where $C = i\gamma_2 \gamma_0$ is the charge-conjugation operator; $\xi_-$ and $\xi_+$ have the color structures $\mathbf{3}_{\bar c \bar s} \otimes \mathbf{\bar 3}_{ud}$ and $\mathbf{\bar 6}_{\bar c \bar s} \otimes \mathbf{6}_{ud}$, respectively; the two matrices $\Gamma_1$ and $\Gamma_2$ can be
\begin{eqnarray}
\Gamma_1 \otimes \Gamma_2 &=& \gamma_5 \otimes \gamma_\mu\gamma_5 \, , \, \gamma_\mu\gamma_5 \otimes \gamma_5 \, ,
\\ \nonumber              && \gamma_\mu \otimes {\bf 1} \, , \, {\bf 1} \otimes \gamma_\mu \, ,
\\ \nonumber              && \gamma^\nu \otimes \sigma_{\mu\nu} \, , \, \sigma_{\mu\nu} \otimes \gamma^\nu \, ,
\\ \nonumber              && \gamma^\nu\gamma_5 \otimes \sigma_{\mu\nu}\gamma_5 \, , \, \sigma_{\mu\nu}\gamma_5 \otimes \gamma^\nu\gamma_5 \, .
\end{eqnarray}
The internal structures described by the former four combinations are more clear, {\it i.e.}, they all contain one $S$-wave diquark/antidiquark field ($q_a^T C \gamma_5 q_b$ of $J^P = 0^+$ or $q_a^T C \gamma_\mu q_b$ of $J^P = 1^+$) and one $P$-wave diquark/antidiquark field ($q_a^T C q_b$ of $J^P = 0^-$ or $q_a^T C \gamma_\mu \gamma_5 q_b$ of $J^P = 1^-$).
We use them to perform QCD sum rule analyses, and find the following one interesting:
\begin{eqnarray}
\xi^\prime_\mu &=& \bar c_a \gamma_5 C \bar s_b^T \left( u_a^T C \gamma_\mu \gamma_5 d_b - u_b^T C \gamma_\mu \gamma_5 d_a \right)
\label{def:xip}
\\ \nonumber && - \bar c_a \gamma_\mu \gamma_5 C \bar s_b^T \left( u_a^T C \gamma_5 d_b - u_b^T C \gamma_5 d_a \right) \, .
\end{eqnarray}
This current has the negative charge-conjugation parity ($C=-$) when containing four identical quarks/antiquarks, and it was used to describe the $Y(2175)$ of $J^{PC} = 1^{--}$ when its quark content is $s s \bar s \bar s$~\cite{Chen:2008ej}.

To better describe the LHCb experiment~\cite{lhcb}, we slightly modify the current $\xi^\prime_\mu$ to be:
\begin{eqnarray}
\xi_\mu(\theta) &=& \cos\theta ~ \bar c_a \gamma_5 C \bar s_b^T \left( u_a^T C \gamma_\mu \gamma_5 d_b - u_b^T C \gamma_\mu \gamma_5 d_a \right)
\label{def:xi}
\\ \nonumber && - \sin\theta ~ \bar c_a \gamma_\mu \gamma_5 C \bar s_b^T \left( u_a^T C \gamma_5 d_b - u_b^T C \gamma_5 d_a \right) \, ,
\end{eqnarray}
where $\theta$ is the mixing angle, and $\xi_\mu(45^{\rm o}) = \xi^\prime_\mu$. We find it well describes the $X_1(2900)$ of $J^P = 1^-$ when the mixing angle is $\theta = 55^{\rm o}$. In the following we use it to perform QCD sum rule analyses.

The current $\xi_\mu$ contains one $S$-wave diquark/antidiquark $q_a^T C \gamma_5 q_b$ of $J^P = 0^+$ and one $P$-wave diquark/antidiquark $q_a^T C \gamma_\mu \gamma_5 q_b$ of $J^P = 1^-$. It has the antisymmetric color structures $\mathbf{3}_{\bar c \bar s} \otimes \mathbf{\bar 3}_{ud}$ and the antisymmetric flavor structure, so its isospin is zero. Accordingly, $\xi_\mu$ would well couple to the $P$-wave $\bar c \bar s u d$ compact tetraquark state of $J^P = 1^-$ and $I=0$, if such a state exists. Assuming the coupling to be ($\epsilon_{\mu}$ is the polarization vector):
\begin{equation}
\langle 0 | \xi_\mu | X_1 \rangle = f_{X_1} \epsilon_\mu \, ,
\label{eq:coupling2}
\end{equation}
we use the current $\xi_\mu$ to perform QCD sum rule analyses. In this case we need to calculate the two-point correlation function
\begin{eqnarray}
\Pi_{\mu\nu}(p^{2}) &=& i\int d^4x ~ e^{ip\cdot x} ~ \langle0| {\bf T} [\xi_\mu(x)\xi_\nu^{\dag}(0)]|0\rangle
\label{equ:Pi2}
\\ \nonumber &=& \left(\frac{p_{\mu}p_{\nu}}{p^2}-g_{\mu\nu}\right)\Pi_1(p^2)+\frac{p_{\mu}p_{\nu}}{p^2}\Pi_0(p^2)\, .
\end{eqnarray}
Since $\Pi_1(p^2)$ receives contributions only from the spin-1 intermediate states, we use it to perform numerical analyses following the same procedures used above.

\begin{figure}[hbt]
\begin{center}
\includegraphics[width=0.45\textwidth]{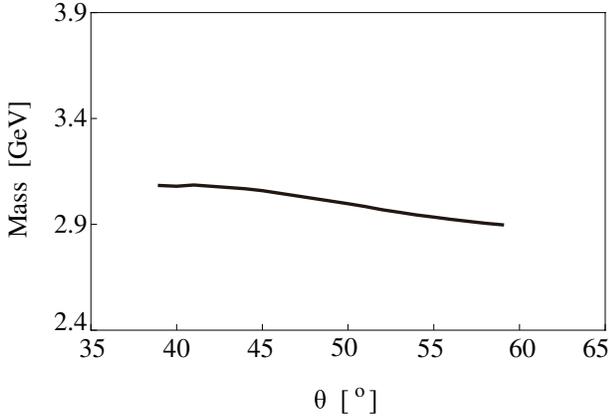}
\caption{
Mass calculated using the current $\xi_\mu(\theta)$ of $J^P=1^-$, as a function of the mixing angle $\theta$. The curve is obtained by setting $s_0 = 10.0$ GeV$^2$, and there exist Borel windows in the region $39^{\rm o } \leq \theta \leq 59^{\rm o}$.}
\label{fig:mixing}
\end{center}
\end{figure}

For the current $\xi_\mu$ of $J^P=1^-$, we show the
extracted mass $m_{X_1}$ with respect to $s_0$ in the left panel of Fig.~\ref{fig:mass}. We show it using the dashed curve, and find that the mass dependence is moderate for
$9.0$ GeV$^2$ $\leq s_0\leq 11.0$ GeV$^2$. This region is just the same as that for the current $\eta$ of $J^P=0^+$. Accordingly, we also fix $s_0=10.0$ GeV$^2$, and extract the Borel window to be $1.79$ GeV$^2$ $\leq M_B^2\leq 1.91$ GeV$^2$. Then we show $m_{X_1}$
with respect to $M_B$ in the right panel of Fig.~\ref{fig:mass} using the dashed curve, and find it very stable inside the Borel window. The hadron mass is evaluated to be
\begin{eqnarray}
m_{X_1,\, 1^-}= 2.93^{+0.13}_{-0.11} \mbox{ GeV}\, .
\label{result:mass1}
\end{eqnarray}
This value is in good agreement with the mass of the $X_1(2900)$ measured by LHCb, suggesting a possible $J^P = 1^-$ $\bar c \bar s u d$ compact tetraquark interpretation for this exotic state. The mass extracted from the current $\xi^\prime_\mu = \xi_\mu(45^{\rm o})$ is slightly larger, that is $3.06^{+0.15}_{-0.10}$~GeV. For completeness, we show the mass dependence on the mixing angle $\theta$ in Fig.~\ref{fig:mixing}.

%
%

Employing the same formalism, we further study its bottom partner, the $P$-wave $\bar b \bar s u d$ compact tetraquark state of $J^P = 1^-$, and extract its mass to be
\begin{equation}
m_{X_{b1},\, 1^-}=6.27^{+0.21}_{-0.22} \mbox{ GeV}\, .
\label{result:massb0}
\end{equation}

$\\$
{\it Summary and discussions.}-----
Very recently the LHCb Collaboration reported their observation of an exotic peak in the $D^- K^+$ invariant mass spectrum. They used two Breit-Wigner resonances to fit this peak, that are the $X_0(2900)$ of $J^P = 0^+$ and the $X_1(2900)$ of $J^P = 1^-$~\cite{lhcb}. This is the first time observing such fully open-flavor tetraquark states, which has opened a new window for studying exotic hadrons.

In this letter we study their possible interpretations using the method of QCD sum rules. We pay special attention to the interesting feature of this experiment that the higher resonance $X_1(2900)$ has a width significantly larger than the lower one $X_0(2900)$, although the $X_1(2900) \to D^-K^+$ decay is $P$-wave and the $X_0(2900) \to D^-K^+$ decay is $S$-wave.

Firstly, we use the meson-meson tetraquark current $\eta$ of $J^P = 0^+$ to perform QCD sum rule analyses, from which the hadron mass is extracted to be
\begin{equation}
m_{X_0,\, 0^+}=2.87^{+0.19}_{-0.14} \mbox{ GeV}\, .
\end{equation}
This value is in good agreement with the mass of the $X_0(2900)$, suggesting a possible $J^P = 0^+$ $D^{*-}K^{*+}$ molecule interpretation for this exotic state. In this picture the width of the $X_0(2900)$ can be naturally explained, {\it i.e.}, it is mainly contributed by the width of the $K^{*+}$ meson.

Secondly, we use the diquark-antidiquark tetraquark current $\xi_\mu$ of $J^P = 1^-$ to perform QCD sum rule analyses, from which the hadron mass is extracted to be
\begin{eqnarray}
m_{X_1,\, 1^-}= 2.93^{+0.13}_{-0.11} \mbox{ GeV}\, .
\end{eqnarray}
This value is in good agreement with the mass of the $X_1(2900)$, suggesting a possible $J^P = 1^-$ $\bar c \bar s u d$ compact tetraquark interpretation for this exotic state. In this picture the width of the $X_1(2900)$ about 110~MeV is also reasonable for a $P$-wave compact tetraquark state.

Since we have not differentiated the $up$ and $down$ quarks in the calculations, the isospin of the $X_0(2900)$ can not be extracted in the present study, and it may contain equally the $\bar D^{*0}K^{*0}$ component. However, the isospin of the $X_1(2900)$ is extracted to be zero ($I = 0$) according to the internal flavor symmetry of the current $\xi_\mu$.

Note that the above two interpretations are just possible explanations, and there exist many other possibilities for the $X_0(2900)$ and $X_1(2900)$~\cite{Liu:2020nil,Huang:2020ptc,Molina:2020hde}. We refer to Ref.~\cite{Karliner:2020vsi} for more discussions, where the authors interpret the $X_0(2900)$ as a $cs \bar u \bar d$ isosinglet compact tetraquark. We also refer to Ref.~\cite{Molina:2010tx}, where the authors predicted a bound $D^* \bar K^*$ state of $J^P=0^+$ within the hidden gauge formalism in a coupled channel unitary approach.

To verify our interpretations, we propose to investigate the $D^{*-} K^{+} \pi^{0}/D^{*-} K^{0} \pi^{+}$ three-body decay channels to confirm the $X_0(2900)$, since its width is mainly contributed by the $K^{*+}$ meson. Besides, we propose to search for their bottom partners in the $B^0 K^+/B^{*0} K^{+} \pi^{0}/B^{*0} K^{0} \pi^{+}$ decay channels, whose masses are predicted to be $m_{X_{b0},\, 0^+}=6.16^{+0.22}_{-0.25}$~GeV and $m_{X_{b1},\, 1^-}=6.27^{+0.21}_{-0.22}$~GeV.

\section*{Acknowledgments}

This project is supported by
the National Natural Science Foundation of China under Grants No.~11722540 and No.~12075019,
the Fundamental Research Funds for the Central Universities,
and the Chinese National Youth Thousand Talents Program.

\end{document}